\documentclass[journal]{IEEEtran}
\usepackage{subfigure,graphicx,booktabs,amssymb,soul,color,graphics,epsfig,mathptmx,times,amsmath,amssymb,color,subfigure,soul,moreverb,ifpdf,cancel,caption,xcolor,amsthm,framed,subfigure,dblfloatfix}

\usepackage{cite}

\usepackage[font=scriptsize]{caption}

\colorlet{shadecolor}{yellow!100}
\ifCLASSINFOpdf
\fi
\begin{document}
\title{Identification of Nonlinear Dynamic Systems Using Type-2 Fuzzy Neural Networks - A Novel Learning Algorithm and a Comparative Study}
\author{Erkan~Kayacan,~\IEEEmembership{Student Member, IEEE,}~Erdal~Kayacan,~\IEEEmembership{Senior Member, IEEE} and Mojtaba Ahmadieh Khanesar,~\IEEEmembership{Member, IEEE}
\thanks{Manuscript received March 28, 2014; revised March 31, 2014 and June 12, 2014; accepted July 14, 2014.}
\thanks{Copyright (c) 2014 IEEE. Personal use of this material is permitted. However, permission to use this material for any other purposes must be obtained from the IEEE by sending a request to pubs-permissions@ieee.org.}
\thanks{Erkan Kayacan is with the Division of Mechatronics, Biostatistics and Sensors, Department of Biosystems, University of Leuven (KU Leuven), Kasteelpark Arenberg 30, B-3001 Leuven, Belgium. e-mail: erkan.kayacan@biw.kuleuven.be}
\thanks{Erdal Kayacan  is with the School of Mechanical \& Aerospace Engineering, Nanyang Technological University, 50 Nanyang Avenue, Singapore 639798. e-mail: {erdal@ntu.edu.sg}}
\thanks{M.A. Khanesar is with the Department of Electrical and Control Engineering, Semnan University, Semnan, Iran. e-mail: ahmadieh@semnan.ac.ir}}
\markboth{\textbf{PREPRINT VERSION: }IEEE TRANSACTIONS ON INDUSTRIAL ELECTRONICS, Volume 62, Issue 3, 2015.}
{Shell \MakeLowercase{\textit{et al.}}: Bare Demo of IEEEtran.cls for Journals}
\maketitle

\begin{abstract}
In order to achieve faster and more robust convergence (especially under noisy working environments), a sliding mode theory-based learning algorithm has been proposed to tune both the premise and consequent parts of type-2 fuzzy neural networks in this paper. Differently from recent studies, where sliding mode control theory-based rules are proposed for only the consequent part of the network, the developed algorithm applies fully sliding mode parameter update rules for both the premise and consequent parts of the type-2 fuzzy neural networks. In addition, the responsible parameter for sharing the contributions of the lower and upper parts of the type-2 fuzzy membership functions is also tuned. Moreover, the learning rate of the network is updated during the online training. The stability of the proposed learning algorithm has been proved by using an appropriate Lyapunov function. Several comparisons have been realized and shown that the proposed algorithm has faster convergence speed than the existing methods such as gradient-based and swarm intelligence-based methods. Moreover, the proposed learning algorithm has a closed form, and it is easier to implement than the other existing methods.
\end{abstract}

\IEEEpeerreviewmaketitle
\section{Introduction}
\IEEEPARstart{I}{nspired} by the central nervous system of human, artificial neural networks (ANNs) are widely known computational tools with their representation capability, even in the case of highly nonlinear and complex structured systems. Fuzzy neural networks (FNNs) combine the capability of fuzzy reasoning to handle uncertain information and the capability of ANNs to learn from input-output data sets in modeling nonlinear dynamic systems. The FNNs have been used in many engineering areas and obtained successful results, such as the identification and control of dynamic systems \cite{1321083,5575419}, temperature control \cite{4813268}, classification \cite{1593641}, energy conversion \cite{4292189}. Another prominent feature of FNNs is that they have also been proven to be universal approximators similar to ANNs \cite{1996}.

There are two different approaches to fuzzy logic system (FLS) design: Type-1 FLSs (T1FLSs) and type-2 FLSs (T2FLSs). The latter is proposed as an extension of the former with the intention of being able to model the uncertainties that invariably exist in the rule base of the system \cite{Mendel_kitap}. Whereas membership functions (MFs) are totally certain in type-1 fuzzy sets, they are themselves fuzzy in type-2 fuzzy sets. The latter case results in a fact that the antecedent and consequent parts of the rules are uncertain. As there are infinite type-1 fuzzy MFs in the footprint of uncertainty of a type-2 fuzzy MF, it is believed that the T2FLSs have the ability of modeling uncertainties in the rule base better than their type-1 counterparts. Therefore, T2FLSs appear to be a more promising method than their type-1 counterparts for handling uncertainties such as noisy data and variable working conditions both in modeling and control purposes \cite{chia09,Juang1,6469210}.

The gradient descent (GD) algorithm is a well-known optimization method to tune the parameters of ANNs and FNNs. However, since the gradient-based algorithms (e.g. dynamic back propagation) include partial derivatives, the convergence speed may be slow especially when the search space is complex. What is more, with the repetitive algorithms, a number of numerical robustness issues may emerge when they are applied over long periods of time \cite{Astrom_Witternmark}. The selection of the learning rate in a GD algorithm is also challenging, because a large value for this parameter may result in instability in the system while a small value may increase the probability of entrapment in a local minima and lessen the speed of convergence. Especially, in FNNs, another issue for a GD algorithm is that obtaining the parameter update rules for the premise part of the network is very complex. In addition to these drawbacks, the tuning process can easily be trapped into a local minimum \cite{Venelinov_1}. For a noisy input-output data set, the performance of gradient-based algorithms might be even worse. In order to improve the performance of these algorithms, several modifications are proposed such as the introduction of momentum term, adaptive learning rate and modification of the traditional GD method by using Levenberg-Marquardt (LM) algorithm \cite{5949558}. As an alternative to GD methods, the use of evolutionary approaches have been suggested \cite{Venelinov_2,5611774}. However, the stability of such approaches is questionable and the optimal values for the stochastic operators are difficult to derive. Furthermore, the computational burden can be very high. To overcome these issues, sliding mode control (SMC) theory-based algorithms are proposed for the parameter update rules of ANNs and type-1 FNNs (T1FNNs) as robust learning algorithms \cite{Parma, Yu}. SMC theory-based learning algorithms cannot only make the overall system more robust but also ensure faster convergence than the traditional learning techniques in online tuning of ANNs and FNNs \cite{Cascella,Efe2000}. Moreover, the parameter update rules are much simpler when compared to other algorithms, such as GD methods. Motivated by the successful results of these learning algorithms on T1FNNs, the derivation of SMC theory-based learning algorithms for the training of type-2 FNNs (T2FNNs) are also proposed \cite{6117080,ACS:ACS1292}.

In order to achieve faster and more robust convergence (especially under noisy working environments), an SMC theory-based learning algorithm has been proposed to tune both the premise and consequent parts of T2FNNs in this paper. Since the proposed learning algorithm does not include any partial derivatives or computationally expensive mathematical operations, e.g. inverse of matrices etc., it is believed that the proposed algorithm is more applicable especially for real-time systems.

In this paper, the major contributions to the T2FNNs are as follows: The first is the proposal of fully SMC theory-based learning rules. It is to be noted that all similar studies in literature consider SMC theory-based rules for only limited number of parameters. For instance, the parameter update rules for the center values of the type-2 fuzzy MFs in \cite{6117080} do not have SMC theory-based rules. The second contribution of this paper is that the proposed algorithm tunes the sharing of the lower and upper MFs in a T2FNN which allows us to manage non-unform uncertainties in the rule base of T2FLSs. As the third contribution, the learning rate is also updated over the simulations.

The body of the paper contains five sections: In Section II, the theoretical basics of T2FLSs are given. The novel SMC theory-based learning rules for the type-2 Gaussian MFs are introduced in Section III. In Section IV, the validation of the proposed novel parameter update rules for the identification of three nonlinear dynamic systems are given. The comparison of different learning techniques is given in Section V. Finally, some conclusions are drawn in Section VI.

\section{Type-2 Fuzzy Logic Systems (T2FLSs)}
\subsection{T2FLSs Overview}
A first-order interval type-2 Takagi-Sugeno-Kang (TSK) fuzzy  \emph{if-then} rule  base with $r$ input variables is preferred in this investigation. Where the consequent parts are crisp numbers, the premise parts are type-2 fuzzy functions. The $r^{th}$ rule is as follows:
\begin{equation*}
R_{r}: \; \text{If} \; x_1 \; \text{is} \;\; \widetilde{A}_{1j} \; \dots \; \text{and} \; x_i \; \text{is} \; \widetilde{A}_{ik}  \;\; \text{and} \dots \; \text{and} \; x_I \; \text{is} \; \widetilde{A}_{Il} \; \text{then}
\end{equation*}
\begin{equation}\label{output_fr}
f_r =  \sum_{i=1}^{I} a_{ri} x_i + b_r
\end{equation}
where $x_i (i=1...I)$ are the inputs of the type-2 TSK model, $\widetilde{A}_{ik}$ is the $k^{th}$ type-2 fuzzy MF $(k=1...K)$ corresponding to the input $i^{th}$ variable, $K$ is the number of MFs for the $i^{th}$ input which can be different for each input.  The parameters $a_r$ and $b_r$ stand for the consequent part and $f_r (r=1...N)$ is the output function.

\begin{figure}[h!]
  \centering
  \includegraphics[width=3in]{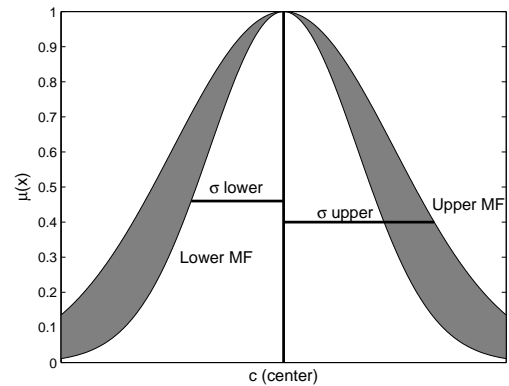}\\
  \caption{Type-2 Gaussian fuzzy MF with uncertain standard deviation}\label{Gaussian}
\end{figure}

The upper and lower type-2 fuzzy Gaussian MFs with an uncertain standard deviation (Fig. \ref{Gaussian}) can be represented as follows:
\begin{equation}
\overline{\mu}_{ik}(x_i)=exp\bigg(-\frac{1}{2}\frac{(x_i-c_{ik})^2}{\overline{\sigma}_{ik}^2} \bigg)
\end{equation}
\begin{equation}
\underline{\mu}_{ik}(x_i)=exp\bigg(-\frac{1}{2}\frac{(x_i-c_{ik})^2}{\underline{\sigma}_{ik}^2} \bigg)
\end{equation}
where $c_{ik}$ is the center value of the $k^{th}$ type-2 fuzzy set for the $i^{th}$ input. The parameters $\overline{\sigma}_{ik}$ and $\underline{\sigma}_{ik}$ are standard deviations for the upper and lower MFs.

\subsection{Interval Type-2 A2-CO TSK Model}

The structure used in this investigation is called A2-C0 fuzzy system \cite{Begian}. In such a structure, first, the lower and upper membership degrees $\underline{\mu }$ and $\overline{\mu }$ are determined for each input signal being fed to the system. Next, the firing strengths of the rules using  the \emph{prod} t-norm operator are calculated as follows:
\begin{equation*}
\underline{w}_r=\underline{\mu}_{\tilde{A}1}(x_1)\ast\underline{\mu}_{\tilde{A}2}(x_2)\ast\cdots\underline{\mu}_{\tilde{A}I}(x_I)
\end{equation*}
\begin{equation} \label{prod}
\overline{w}_r=\overline{\mu}_{\tilde{A}1}(x_1)\ast\overline{\mu}_{\tilde{A}2}(x_2)\ast\cdots\overline{\mu}_{\tilde{A}I}(x_I)
\end{equation}

The consequent part corresponding to each fuzzy rule is a linear combination of the inputs $x_1,\,x_2\,...\,x_I$. This linear function is called $f_r$ and is defined as in (1). The output of the network is calculated as follows:
\begin{equation}\label{Rahib2_6} y_N=q\sum_{r=1}^{N}f_r\underline{\widetilde{w}}_r + (1-q)\sum_{r=1}^{N}f_r\widetilde{\overline{w}}_r
\end{equation}


\noindent where $\widetilde{\underline{w_{r}}}$ and $\widetilde{\overline{w_{r}}}$ are the normalized values of the lower and the upper output signals from the second hidden layer of the network as follows:
\begin{eqnarray}\label{wtilde}
\widetilde{\underline{{w_r}}} = \frac{\underline{w_r}}{\sum_{i=1}^{N}\underline{w_r}} \mbox{ and }
\widetilde{\overline{{w_r}}} = \frac{\overline{w_r}}{\sum_{i=1}^{N}\overline{w_r}}
\end{eqnarray}

The design parameter, $q$, weights the sharing of the lower and the upper firing levels of each fired rule.  This parameter can be a constant (equal to $0.5$ in most cases) or a time varying parameter. In this investigation, the latter is preferred. In other words, the parameter update rules and the proof of the stability of the learning process are given for the case of a time varying $q$.

The following vectors can be specified:

\noindent $\widetilde{\underline{{W}}}\left(t\right)=\left[\widetilde{\underline{{w_{1}}}} \left(t\right)\; \widetilde{\underline{{w_{2}}}} \left(t\right)...\; \widetilde{\underline{{w_{N}}}} \left(t\right)\right]^{T} $,

\noindent $\widetilde{\overline{{W}}}\left(t\right)=\left[\widetilde{\overline{{w_{1}}}} \left(t\right)\; \widetilde{\overline{{w_{2}}}} \left(t\right)...\; \widetilde{\overline{{w_{N}}}} \left(t\right)\right]^{T} $ and $F=[f_{1} \; f_{2} \; ...f_{N}] $

The following assumptions have been used in this investigation: The time derivative of both the input signals and output signal can be considered bounded:
\begin{equation}
|\dot{x}_i(t)|\leq B_{\dot{x}}, \;\; min (x^2_i(t)) =  B_{x^2}, \hspace{0.15cm}(i=1\ldots I) \hspace{0.15cm} \text{and} \hspace{0.15cm} |\dot{y}(t)|\leq B_{\dot{y}} \hspace{0.15cm} \forall t
\end{equation}
where $B_{\dot{x}}$, $B_{x^2}$ and $B_{\dot{y}}$ are assumed to be some known positive constants.

\section{Sliding Mode Control Theory-based Learning Algorithm}
The zero value of the learning error coordinate can be defined as a time-varying sliding surface in (\ref{slidingsurface}). The condition defined in (\ref{slidingsurface}) guarantees that when the system is on the sliding surface, the output of the network, $y_N(t)$, will perfectly follow the desired output signal, $y(t)$, for all time $t>t_h$. The time instant $t_h $ is defined to be the hitting time for being $e(t)=0$.
\begin{equation} \label{slidingsurface}
S\big(e(t)\big)=e(t)=y_N(t)-y(t)=0
\end{equation}

\textit{Definition:} A sliding motion will appear on the sliding manifold $S\left(e(t)\right)=e(t)=0$
after a time $t_{h}$, if the condition $S(t)\dot{S}(t)<0$ is satisfied for all
$t$ in some nontrivial semi-open subinterval of time of the form $\left[t,t_{h} \right)\subset \left(0, t_{h} \right)$.

It is desired to devise a dynamical feedback adaptation mechanism, or an online learning algorithm for the parameters of the T2FNN considered, such that the sliding mode condition of the above definition is enforced.

\subsection{The Proposed Parameter Update Rules for the T2FNN}
The parameter update rules for the T2FNN proposed in this paper are given by the following theorem.

\emph{Theorem 1:} If the adaptation laws for the parameters of the considered T2FNN are chosen as:
\begin{equation} \label{c_ik}
\dot{c}_{ik} = \dot{x}_{i} + (x_{i} - c_{ik}) \alpha_1 \textrm{sgn}\left(e \right)
\end{equation}

\begin{equation}\label{sigma_1i_lower}
\dot{\underline{\sigma}}_{ik} = - \bigg( \underline{\sigma}_{ik} + \frac{ (\underline{\sigma}_{ik} )^3}{(x_{i} - c_{ik})^2} \bigg) \alpha_1  \textrm{sgn}\left(e \right)
\end{equation}
\begin{equation}\label{sigma_1i_upper}
\dot{\overline{\sigma}}_{ik} = - \bigg( \overline{\sigma}_{ik} + \frac{ (\overline{\sigma}_{ik} )^3}{(x_{i} - c_{ik})^2} \bigg) \alpha_1  \textrm{sgn}\left(e \right)
\end{equation}
\begin{equation} \label{ar}
\dot{a}_{ri} =-x_{i}\frac{q \tilde{\underline{w}}_{r} + (1-q) \tilde{\overline{w}}_{r}}{(q \tilde{\underline{w}}_{r} + (1-q) \tilde{\overline{w}}_{r} )^{T} (q \tilde{\underline{w}}_{r} + (1-q) \tilde{\overline{w}}_{r})}\alpha \textrm{sgn}\left(e \right)
\end{equation}
\begin{equation} \label{br}
\dot{b}_{r} = -\frac{q \tilde{\underline{w}}_{r} + (1-q) \tilde{\overline{w}}_{r}}{(q \tilde{\underline{w}}_{r} + (1-q) \tilde{\overline{w}}_{r} )^{T} (q \tilde{\underline{w}}_{r} + (1-q) \tilde{\overline{w}}_{r})}\alpha \textrm{sgn}\left(e \right)
\end{equation}
\begin{equation}\label{dotq}
\dot{q} =-\frac{1}{F(\widetilde{\underline{W}}-\widetilde{\overline{W}})^{T}}\alpha sgn(e)
\end{equation}
where $\alpha$ is an adaptive learning rate with an adaptation law as follows:
\begin{equation} \label{alpha}
\dot \alpha=\gamma (I+2)\mid e\mid-\nu\gamma \alpha,\,\,\,0<\gamma,\nu
\end{equation}

Then, given an arbitrary initial condition $e(0)$, the learning error $e(t)$ will converge  to zero within a finite time $t_h$.

\textit{Proof: }The reader is referred to  Appendix A.

\textbf{Remark 1.} It is to be noted that in \eqref{alpha}, the parameter $\gamma$  has a small positive real value which is interpreted as the learning rate for the adaptive learning rate. Moreover, the first term of the adaptation law of \eqref{alpha} is always positive which may result in unnecessarily large value for $\alpha$. In order to avoid bursting in the parameter $\alpha$, the second term introduces a reset mechanism which avoids a possible parameter bursting in $\alpha$. The value of $\nu$ should be selected very small to avoid it from interrupting the adaptation mechanism.

\textbf{Remark 2.} As can be seen from \eqref{alpha}, the learning rate considered in this paper is itself adaptive and does not need to be known as a priori. This is an obvious superiority of the current approach with respect to previous approaches in which the upper bounds of the states of the system should be known as a priori in order to choose an appropriate value for the learning rate.

In order to avoid division by zero in the adaptation laws of \eqref{c_ik}-\eqref{dotq} an instruction is included in the algorithm to make the denominator equal to $0.001$ when its calculated value is smaller than this threshold.

It is well-known that sliding mode control suffers from high-frequency oscillations in the control input, which are called \emph{chattering}. The following are the two common methods used to eliminate chattering \cite{Slotine}:
\begin{enumerate}
  \item Using a saturation function to replace the signum function.
  \item Inserting a boundary layer so that an equivalent control replaces the corrective one when the system is inside this layer.
\end{enumerate}

In order to reduce the chattering effect, the following function is used in this paper with $\delta_s=0.05$ instead of the signum function in the dynamic strategy:
\begin{equation}\label{chatter}
\textrm{sgn}(e):=\frac{e}{|e|+\delta_s}
\end{equation}

\textbf{Remark 3.} Since the output of the T2FNN is quite sensitive to the changes in the parameters of the antecedent parts, different values for the learning rates of the antecedent and consequent part parameters are used. In other words, a smaller value ($\alpha_1$) is chosen for the antecedent parts.

\section{Simulation Studies}
\begin{figure}[h!]
\centering
\includegraphics[width=3in]{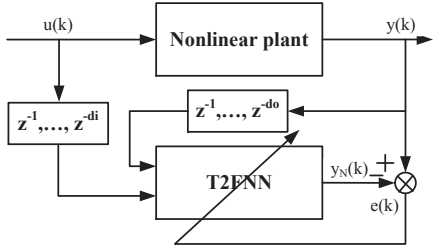}\\
\caption{Identification scheme}
\label{fig3}
\end{figure}

The identification problem involves the finding of the input-output relations of the system. In Fig. \ref{fig3} the structure of the identification scheme is shown. The inputs to the T2FNN based identifier are the external input signals, its one-, two, \ldots, $d_i$- step delayed values and the one-, two-, \ldots, $d_o$- step delayed outputs of the plant. The problem is defined as to find such values of the parameters of the T2FNN structure that the difference between the plant output $y(k)$ and the identifier output $y_N(k)$ will be minimum for all input values of $u(k)$.

The number of the parameters to be updated in the T2FNN structure is the summation of the following parameters:
\begin{itemize}
  \item The number of the parameters for the premise parts of the rules (the center and sigma values of the Gaussian MFs)
  \item The number of the parameters for the consequent parts of the rules ($a_{ri}$ and $b_r$ matrices in \eqref{output_fr})
  \item The parameter $q$ (in \eqref{Rahib2_6})
   \item The learning rate $\alpha$
\end{itemize}

In order to evaluate the identification performance of the proposed learning algorithm, number of simulation studies are carried out. The systems are taken from literature in order to be able to make a fair comparison with the existing algorithms.

As a performance criterion, the root-mean-square-error (RMSE) given in \eqref{rmse} is used:
\begin{equation}\label{rmse}
RMSE=\sqrt{\frac{\sum_{k=1}^K (y(k)-y_N(k))^2}{K}}
\end{equation}
where $K$ is the number of samples.

The identification performance of the proposed learning algorithm has been compared with GD, particle swarm optimization (PSO), SMC theory-based online learning for T1FNNs, and the extended sliding mode on-line algorithm for T2FNN presented in \cite{Topalovv}. It is to be noted that even if the network structure is the same with the one in \cite{Topalovv}, the proposed learning rules in this investigation are completely novel and fully sliding mode. In all the examples in this section, the network is designed with three inputs and one output. The inputs are the input signal to the plant, the two delayed signals from the plant output with a discretization period $T_o$ of 1ms. Each input is fuzzified by using three Gaussian type-2 fuzzy MFs with a fixed center and uncertain standard deviation. To be able to make a fair comparison, each experiment has been realized for ten times with a random initialization of the network parameters, and the average numbers are given in the paper.

\subsection{Example 1: Identification of a non-BIBO nonlinear plant} The proposed identification procedure is applied to a non-bounded-input-bounded-output (non-BIBO) nonlinear plant model \cite{363441} described by the following equation:

\begin{eqnarray}\label{non-BIBO}
y(k+1)&=&0.2y^2(k)+0.2y(k-1) \nonumber \\
     &+&0.4sin\Big(0.5\big(y(k)+y(k-1)\big)\Big)\nonumber \\
     &&\times   \hspace{0.1cm} cos\Big(0.5\big(y(k)+y(k-1)\big)\Big) \nonumber \\
     &+&1.2u(k)
\end{eqnarray}

The plant output may diverge for a sequence of uniformly bounded input signals. For instance, when a step input $u(k)=0.83$ is applied to the system, the output of the system diverges. On the other hand, when a step input $u(k)<0.83$ is applied to the system, the possible maximum output of the system is approximately equal to 2.26. Thus, the input signal in this investigation has the following form \cite{Topalovv}:
\begin{equation}\label{u_k}
u(k)=0.5e^{-0.1kT_o}sin(5kT_o)
\end{equation}

\begin{figure}[htb]
\centering
\subfigure[ ]{
\includegraphics[width=1.5in]{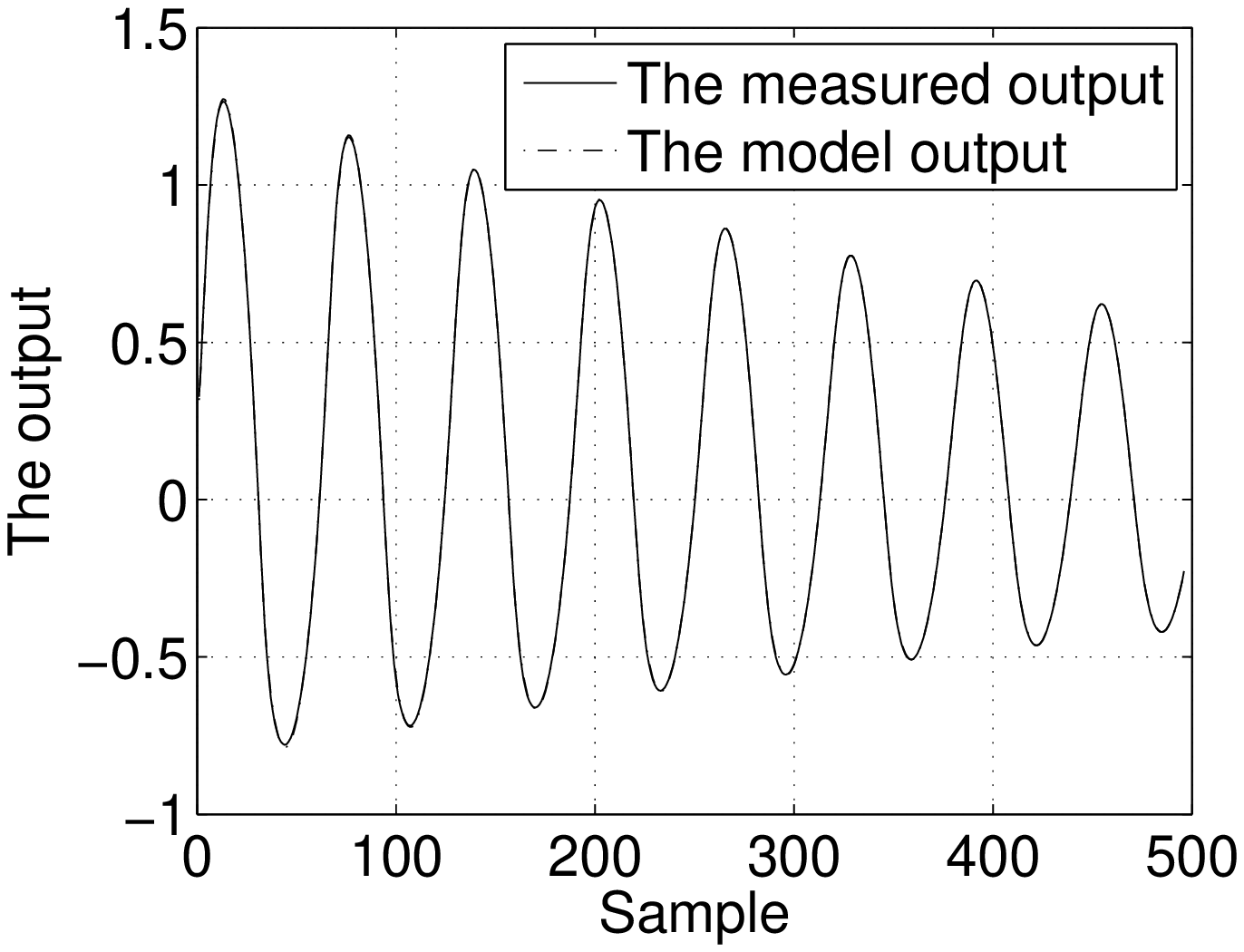}
\label{target_and_model}
}
\subfigure[ ]{
\includegraphics[width=1.5in]{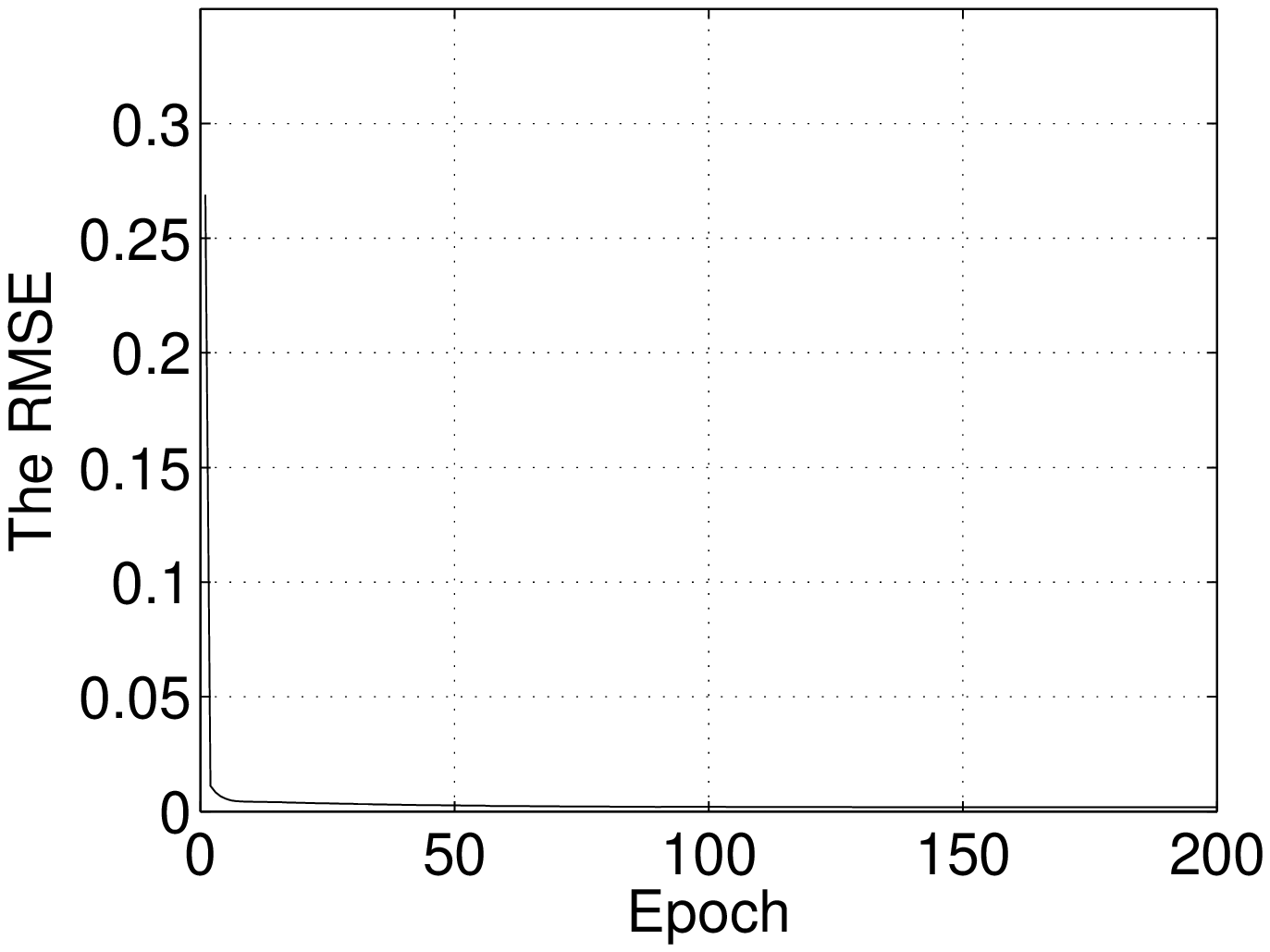}
\label{RMSE}
}
\subfigure[ ]{
\includegraphics[width=1.5in]{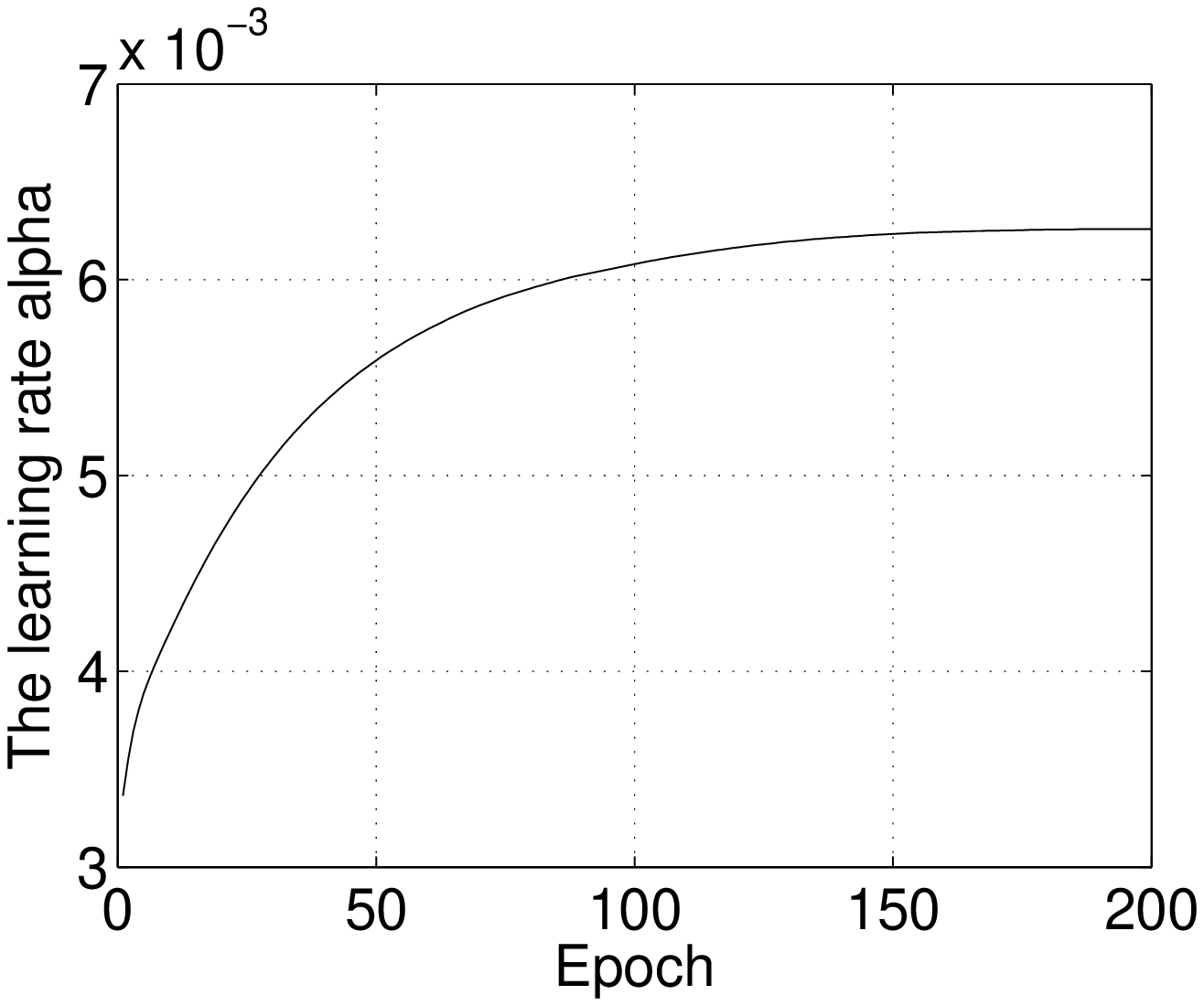}
\label{aalpha}
}
\subfigure[ ]{
\includegraphics[width=1.5in]{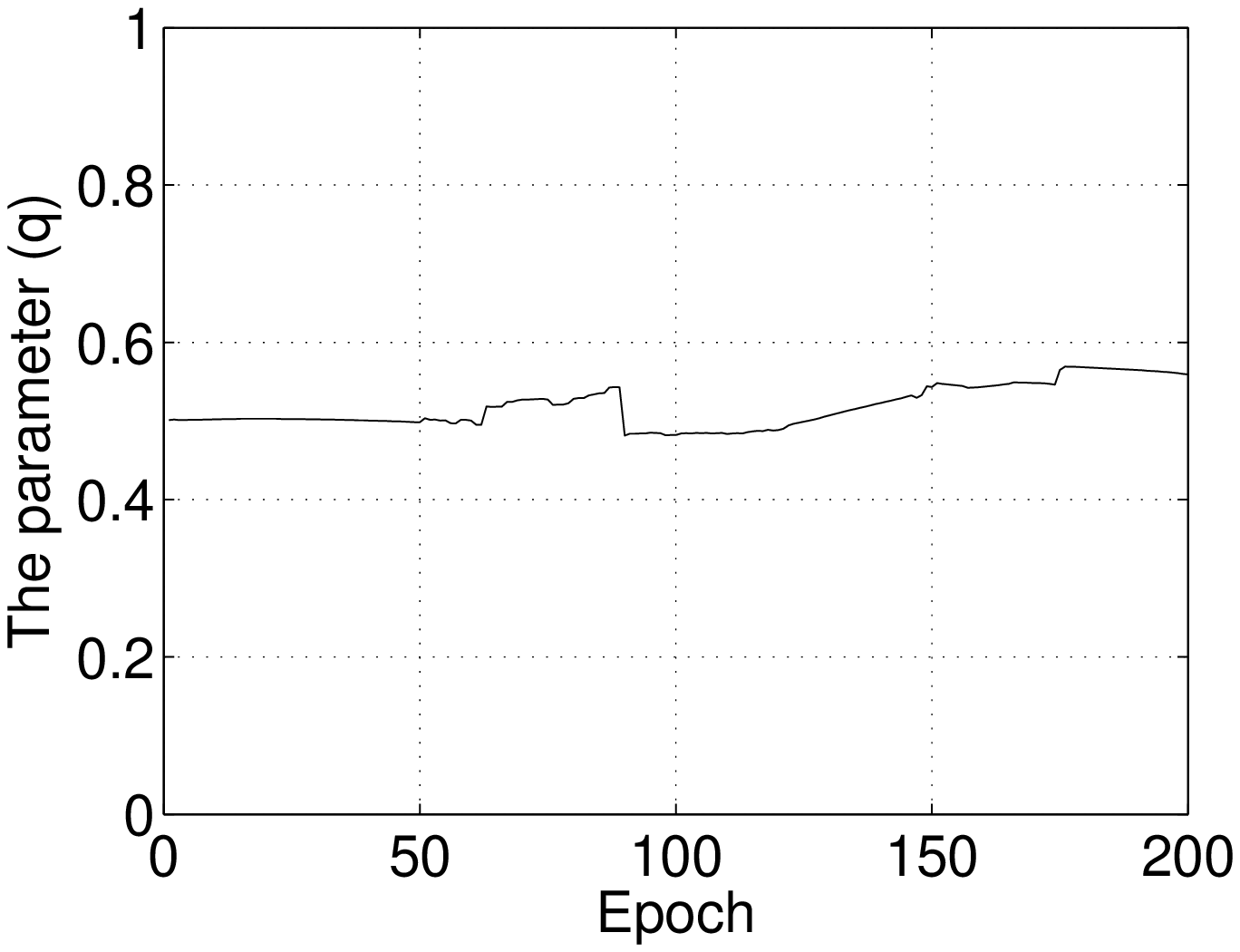}
\label{The_parameter_q}
}
\caption[Optional caption for list of figures]{The output of the model and the T2FNN system (a) RMSE versus epoch number (b) The adaptation of the learning rate (c) The adaptation of the parameter q (d)}
\label{Ex1}
\end{figure}

Whereas Fig. \ref{target_and_model} demonstrates the output of the model and the real-time system, Fig. \ref{RMSE} shows the RMSE values versus epoch number which indicates a stable learning with the proposed learning algorithm. As can be seen from Fig. \ref{target_and_model}, the T2FNN gives accurate modeling results. Thanks to the novel fully sliding mode parameter update rules in this paper, the presented results  are significantly better when compared to the ones in \cite{Topalovv}. In Fig. \ref{aalpha}, the adaptation of the learning rate is presented. Thanks to the reset mechanism presented in \eqref{alpha} the learning rate does not go to infinity; it converges to an appropriate value. In Fig. \ref{The_parameter_q}, the adaptation of the parameter $q$ is presented which is also learnt by the proposed algorithm. By doing so, the contributions of the upper and lower MFs are also tuned during the simulations.

\subsection{Example 2: Identification of a second-order nonlinear time-varying plant} In the second example, the proposed identification procedure is applied to a  second-order nonlinear time-varying plant \cite{Abiyev1} described by the following equation:
\begin{equation}\label{non-time-varying}
y(k)=\frac{x_1x_2+x_3}{x_4}
\end{equation}
where $x_1=y(k-1)y(k-2)y(k-3)u(k-1)$, $x_2=y(k-3)-b(k)$, $x_3=c(k)u(k)$ and  $x_4=a(k)+y(k-2)^2+y(k-3)^2$.

The time-varying parameters $a$, $b$ and $c$ in \eqref{non-time-varying} are given by the following equation:
\begin{eqnarray}\label{time-varying-parameters}
a(k)&=&1.2-0.2cos(2\pi k/T)  \nonumber \\
b(k)&=&1-0.4sin(2\pi k/T)  \nonumber \\
c(k)&=&1+0.4sin(2\pi k/T)
\end{eqnarray}
where $T=1000$ is the time span of the test.

There exist number of papers in the literature that claim that the performance of T2FNNs is better than their type-1 counterparts under uncertain working conditions and the claim is tried to be justified by simulation studies only for some specific systems. However, this claim is justified numerically in a general way in \cite{Khanesar}. Since the system described in \eqref{non-time-varying} is a nonlinear and time-varying system which has to operate under uncertain conditions. Therefore, this system is considered to be an appropriate  system to show the performance of the T2FNN which is trained with the novel parameter update rules in this paper.

\begin{figure}[htb]
\centering
\subfigure[ ]{
\includegraphics[width=1.5in]{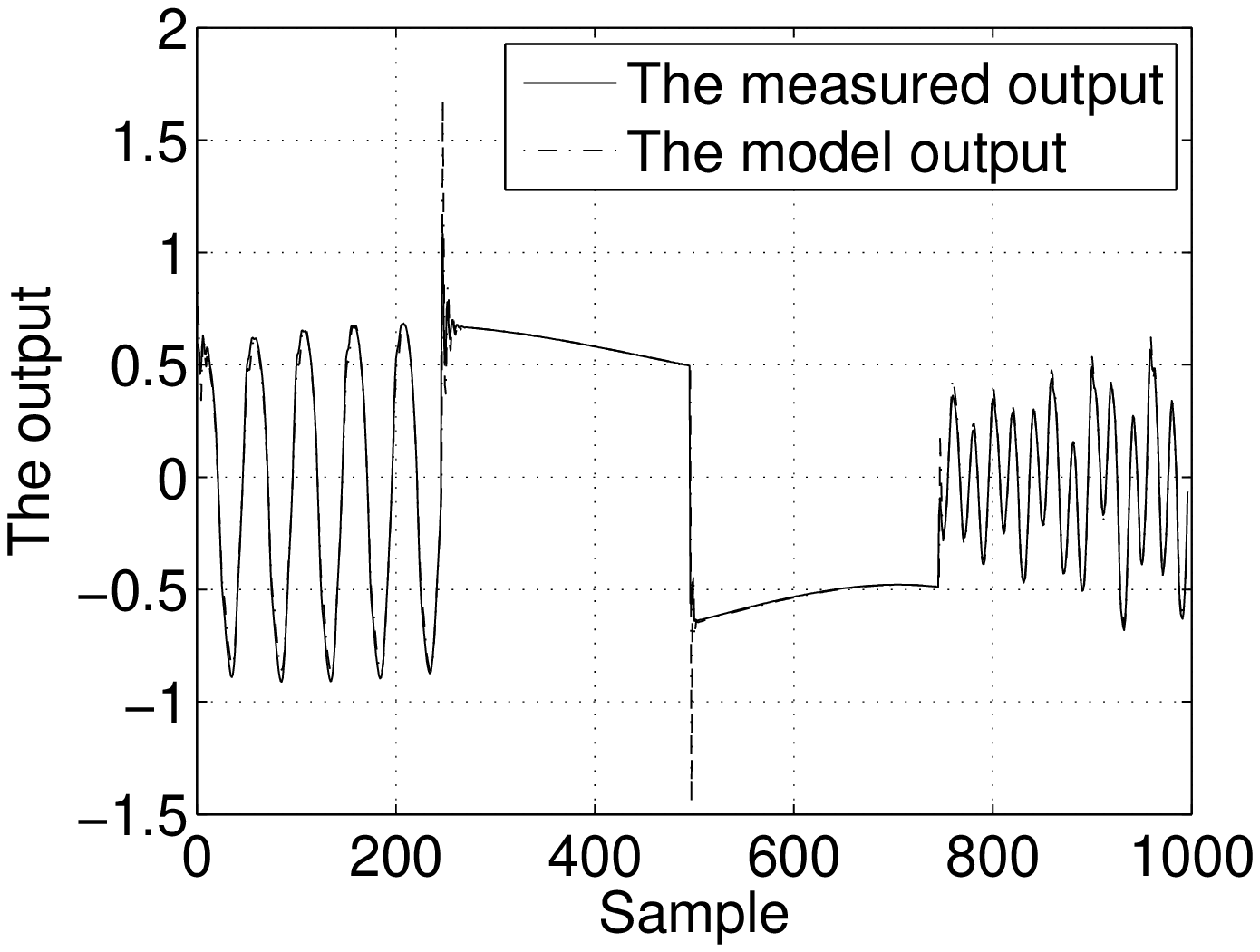}
\label{target_and_model11}
}
\subfigure[ ]{
\includegraphics[width=1.5in]{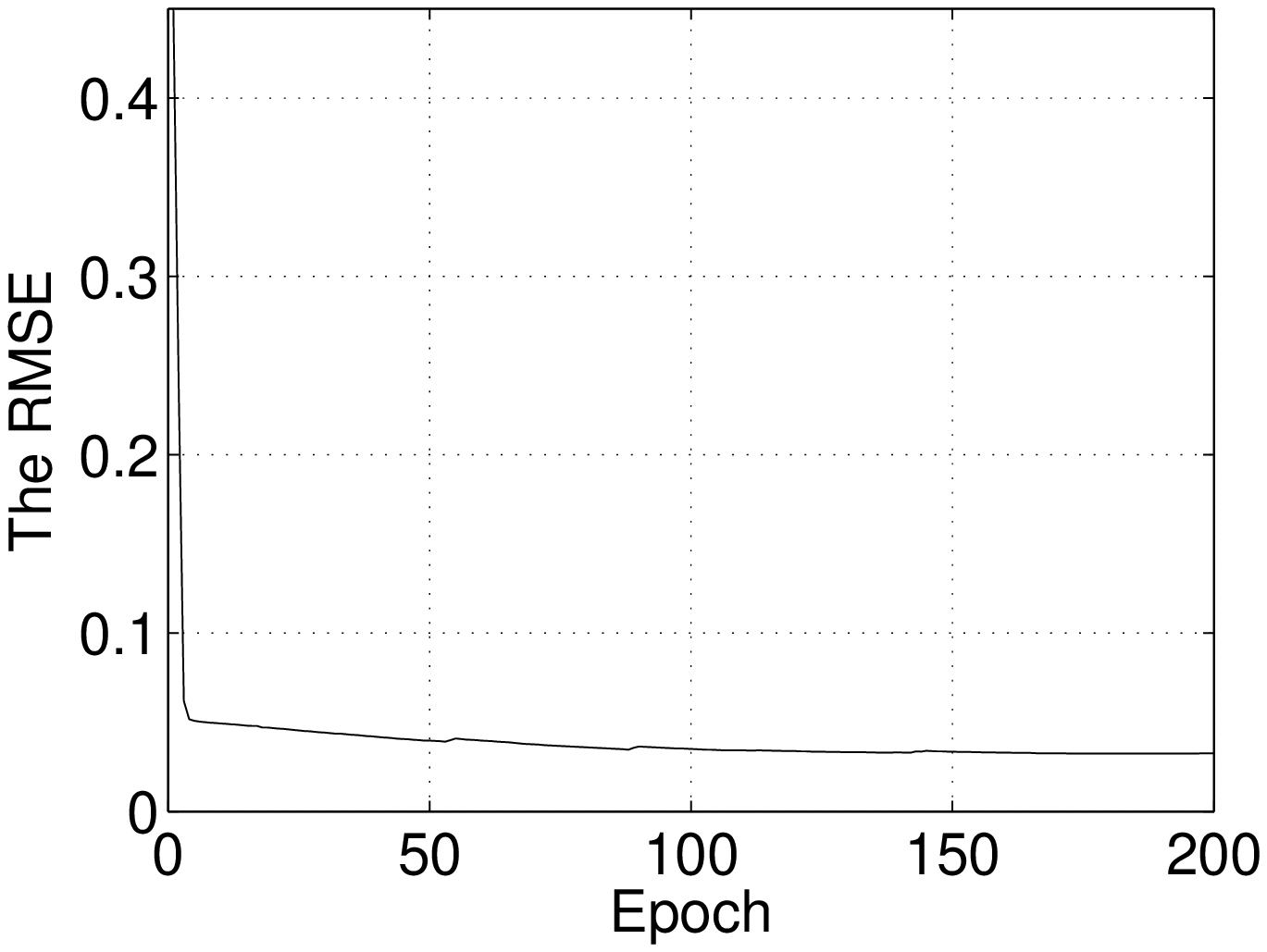}
\label{RMSE11}
}
\subfigure[ ]{
\includegraphics[width=1.5in]{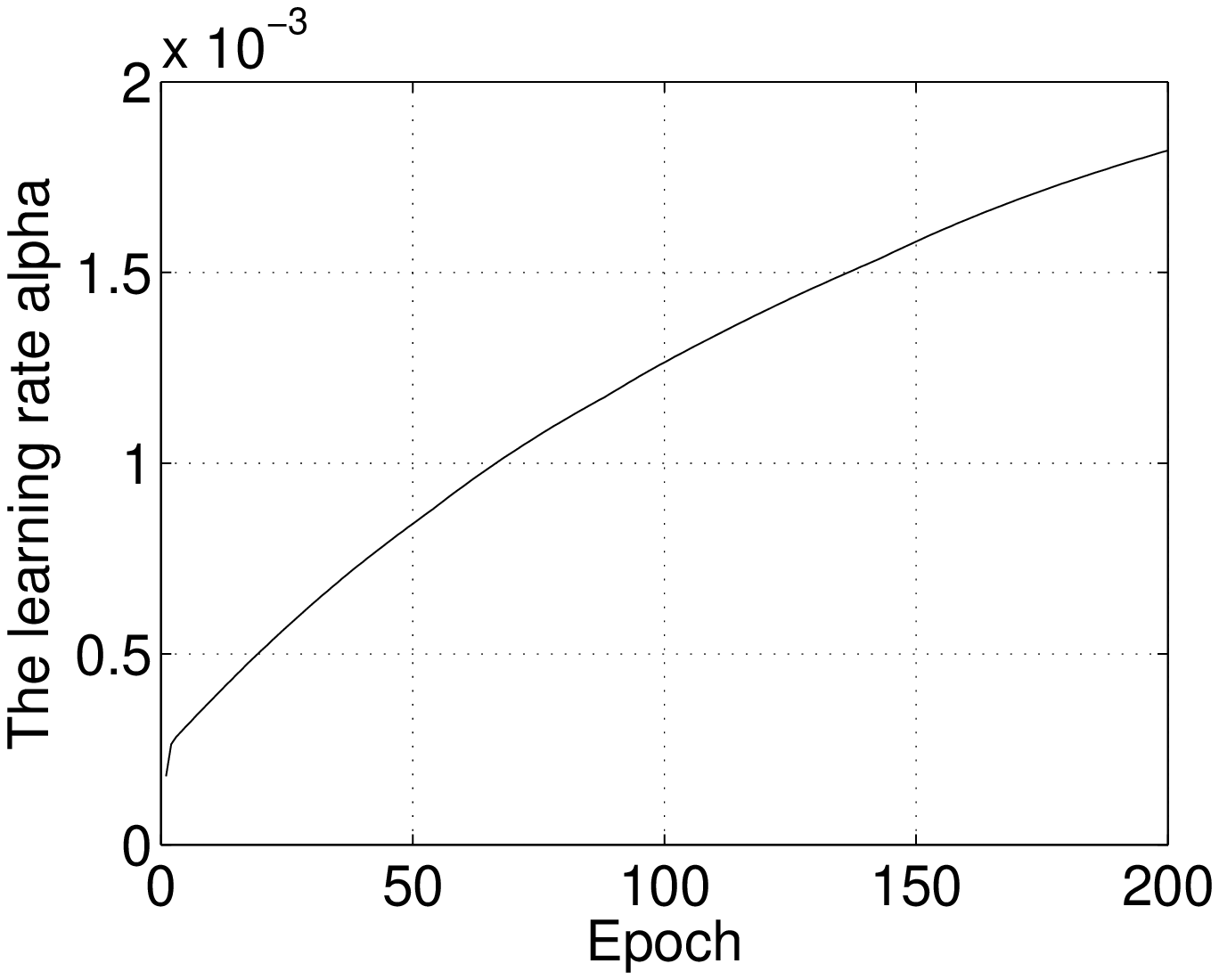}
\label{aalpha11}
}
\subfigure[ ]{
\includegraphics[width=1.5in]{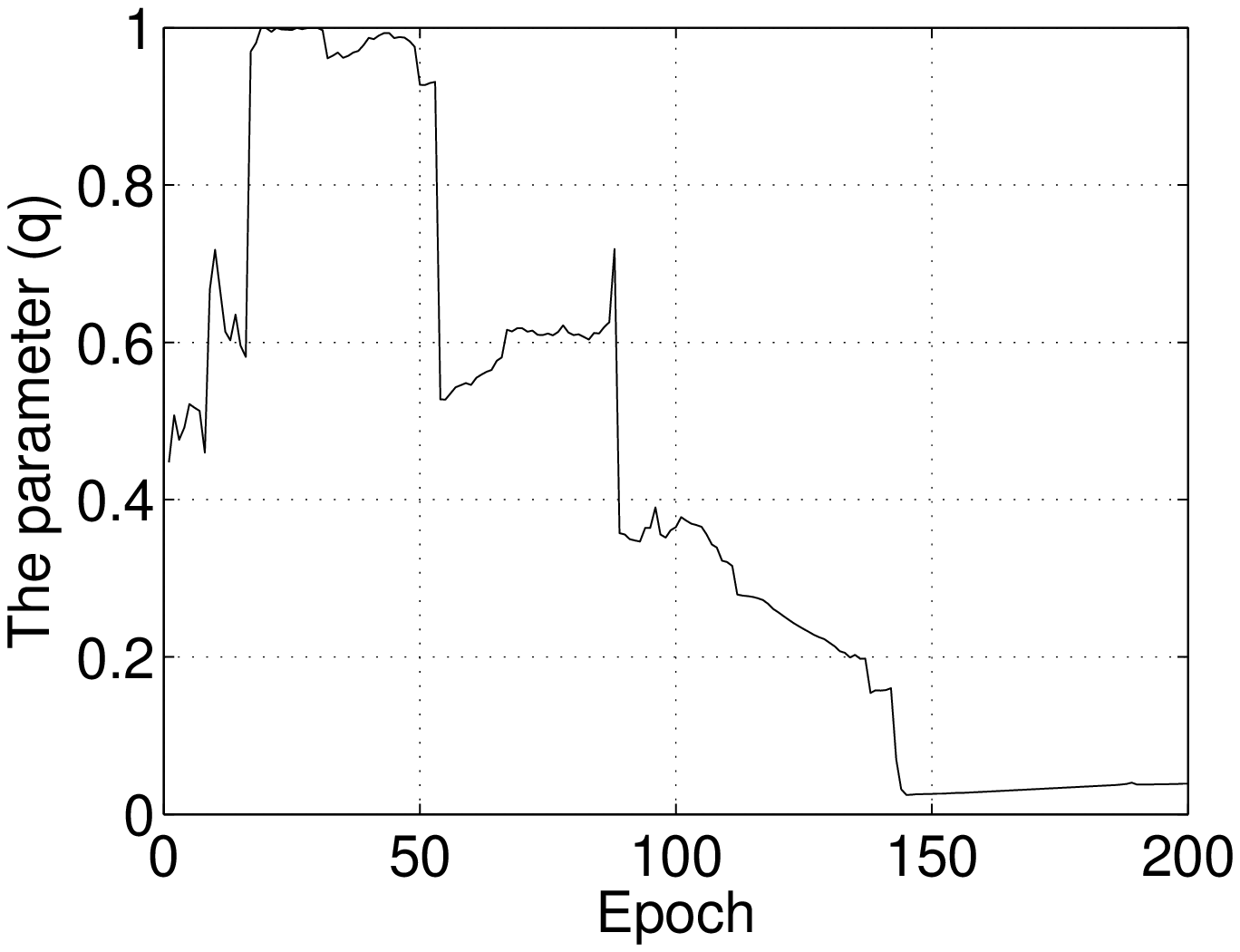}
\label{The_parameter_q11}
}
\label{Ex2}
\caption[Optional caption for list of figures]{The output of the model and the T2FNN system (a) RMSE versus epoch number (b) The adaptation of the learning rate (c) The adaptation of the parameter q (d)}
\end{figure}

Similar to Fig. \ref{Ex1}, whereas Fig. \ref{target_and_model11} demonstrates the output of the model and the real-time system, Fig. \ref{RMSE11} shows the RMSE values versus epoch number which indicates a stable learning with the proposed learning algorithm. Thanks to the novel fully sliding mode parameter update rules in this paper, the presented results  are quite similar to the ones in \cite{Abiyev1}. However, it is to be noted that the paper \cite{Abiyev1}  uses more complex fuzzy logic system structure and more complex parameter adaptation rules when compared to the ones in this investigation. Moreover, the adaptation laws in \cite{Abiyev1} do not have closed (explicit) forms. In Fig. \ref{aalpha11}, the adaptation of the learning rate is presented. Thanks to the reset mechanism presented in \eqref{alpha} the learning rate does not go to infinity; it converges to an appropriate value. In Fig. \ref{The_parameter_q11}, the adaptation of the parameter $q$ is presented which is also tuned by the proposed algorithm. By doing so, the contributions of the upper and lower MFs are also tuned during the simulations.

\section{Analysis and Discussion}
In Table \ref{Table1}, the comparison of different learning techniques is given with respect to their identification performance and computation time. It can be seen that the identification performances of extended Kalman filter and the proposed SMC theory-based learning algorithm are similar to each other, and they seem to be the best when compared to other techniques. However, the computation time of the proposed SMC theory-based learning algorithm is significantly lower than the other methods. This conclusion results in a fact that the proposed method in this paper is more practical in real-time applications.

The reason for having the largest computation time for the extended Kalman filter is that this algorithm includes the manipulation of some high order matrices. A large amount of memory is used by these high dimensional matrices which makes the algorithm difficult to implement in most of the real time applications. On the other hand, there are no matrix manipulations in the proposed SMC theory-based rules. Moreover, GD, LM, EKF and other gradient-based training methods include the calculations of the partial derivatives of the output with respect to the parameters which is very difficult, and do not have any closed form. On the other hand, as can be seen from the adaptation laws proposed in this paper, the parameter update algorithm has closed form.

In order to have a better comparison, the proposed method is compared with an algorithm which is composed of particle swarm optimization (PSO) and GD. This algorithm has been previously used to train recurrent type-1 fuzzy neural networks \cite{khanesar2007hybrid} and interval T2FNNs \cite{5611774}. The reader may refer to these references for a complete description about the training algorithm. The basic idea behind of this algorithm is that the parts which appear nonlinearly in the output can be trained by using PSO while the parameters which appear linearly in the output can be trained using GD. This is because its computational back-bone which makes it more logical than random optimization methods. This method is also implemented on A2-C0 fuzzy model which is considered in this paper and the results are summarized in Table I. As can be seen from the table, the results obtained using this method are quite close to those of SMC theory-based training algorithm while the computational time is much higher. The reason behind this computational burden is that the PSO+GD method necessitates too many feed-forward computation of T2FNN which is very complex and time consuming task.

\begin {table}[h!]
\begin{center}
\caption {Comparison of different learning techniques} \label{Table1}
\begin{tabular}{cccc}
\toprule
\multicolumn{3}{c}{\hspace{3cm}Performance} \\
\cmidrule(r){2-3}
& Training    & Testing & Computation time (s) \\
\midrule
SMC-based learning       &0.0342  &0.0280  &85.3026   \\
GD       &0.0636  &0.0919  &120.5814  \\
LM    &0.0463  &0.0567  &296.3332  \\
EKF &0.0276  &0.0266  &218.7358  \\
PSO+GD &0.0307  &0.0525  &7536.400  \\
\bottomrule
\label{Table1}
\end{tabular}
\end{center}
\end {table}

One issue that should be taken into account is that the adaptation laws proposed in this paper are continuous. However, for the simulation of the method in a computer, an optimal sampling time should be chosen. The choice of the optimal sampling time may be a problem, because a very large value for the sampling time may cause instability in the system.

\section{Conclusions and Future Work}

\subsection{Conclusions}
A novel fully sliding mode parameter update rules have been proposed for the training of interval T2FNNs for the identification of nonlinear and time-varying dynamic systems. The performance of the learning algorithm has been tested on the identification of two nonlinear systems which are taken from literature to be able to make a fair comparison. The simulation results indicate the potential of the proposed structure in real time systems since the computation time of the proposed algorithm is significantly lower than the other methods as well as it gives high identification accuracy. It is to be noted that these parameter update rules can also be used for the control purposes in which the computation time is prominent.

One issue which should be taken into account is that the adaptation laws  proposed in this paper are continous. However, for the simulation of the method in a computer, an optimal sampling time should be chosen. The choice of the optimal sampling time may be a problem, because a very large value for the sampling time may cause an instability in the system.

\subsection{Future Work}
The proposed method with its stability analysis is valid for only  the type-2 Gaussian membership functions with uncertain standard deviation. As a future work, the extension of the proposed method to other types of membership functions like Gaussian type-2 membership functions with uncertain centers, Elliptic type-2 MFs and so on are some interesting topics to be investigated.

\appendices
\section{Proof of Theorem 1}
The time derivative of \eqref{wtilde} is calculated as follows:
\begin{equation} \label{w_r}
\dot{\underline{\tilde{w}}}_{r} = - \underline{\tilde{w}}_{r} \underline{K}_{r} + \underline{\tilde{w}}_{r} \sum_{r=1}^{N} \underline{\tilde{w}}_{r}  \underline{K}_{r}; \;\;
\dot{\tilde{\overline{w}}}_{r} = - \tilde{\overline{w}}_{r} \overline{K}_{r} + \tilde{\overline{w}}_{r} \sum_{r=1}^{N} \tilde{\overline{w}}_{r}  \overline{K}_{r}
\end{equation}
where
\begin{equation*} \label{A_ik_lower}
\underline{A}_{ik} = \frac{ x_{i} - c_{ik} } { \underline{\sigma}_{ik} } \;\; \text{and} \;\; \overline{A}_{ik} = \frac{ x_{i} - c_{ik} } { \overline{\sigma}_{ik} }
\end{equation*}
\begin{equation*} \label{K_r_lower}
\underline{K}_{r} = \sum_{i=1}^{I}  \underline{A}_{ik} \dot{\underline{A}}_{ik} \;\; \text{and} \;\;  \overline{K}_{r} = \sum_{i=1}^{I}  \overline{A}_{ik} \dot{\overline{A}}_{ik}
\end{equation*}

If \eqref{c_ik}-\eqref{sigma_1i_upper} are inserted into the equations above, \eqref{K_r} can be obtained:
\begin{equation} \label{K_r}
\overline{K}_{r} = \underline{K}_{r} = \sum_{i=1}^{I}  \overline{A}_{ik} \dot{\overline{A}}_{ik} = \sum_{i=1}^{I}  \underline{A}_{ik} \dot{\underline{A}}_{ik} = I \alpha \textrm{sgn}\left(e \right)
\end{equation}

By using the following Lyapunov function, the stability condition is checked as follows:
\begin{equation} \label{V}
V=\frac{1}{2} e^2 + \frac{1}{2\gamma}(\alpha-\alpha^*)^2,\,\,\,\,0<\gamma
\end{equation}

The time derivative of \eqref{V} can be calculated as follows:
\begin{equation} \label{dotV}
\dot{V}=\dot{e} e + \frac{1}{\gamma}\dot{\alpha}(\alpha-\alpha^*) = e (\dot{y}_N - \dot{y}) + \frac{1}{\gamma}\dot{\alpha}(\alpha-\alpha^*)
\end{equation}

Differentiating \eqref{Rahib2_6}, the following term can be obtained:
\begin{eqnarray} \label{dotyn}
\dot{y}_N & = & \dot{q} \sum_{r=1}^{N} f_r \tilde{\underline{w}}_{r} + q \sum_{r=1}^{N} ( \dot{f}_r \tilde{\underline{w}}_{r} + f_r \dot{\tilde{\underline{w}}}_{r} ) - \dot{q} \sum_{r=1}^{N} f_r \tilde{\overline{w}}_{r} \nonumber \\
&& + (1 - q ) \sum_{r=1}^{N} ( \dot{f}_r \tilde{\overline{w}}_{r} + f_r \dot{\tilde{\overline{w}}}_{r} )
\end{eqnarray}

By using \eqref{w_r}, \eqref{K_r} and \eqref{dotyn}, the following term can be obtained:
\begin{eqnarray} \label{dotyn2}
\dot{y}_N & = & \dot{q} \sum_{r=1}^{N} f_r \tilde{\underline{w}}_{r} + q \sum_{r=1}^{N} \Big( \dot{f}_r \tilde{\underline{w}}_{r} + f_r (- \underline{\tilde{w}}_{r} \underline{K}_{r} + \underline{\tilde{w}}_{r} \sum_{r=1}^{N} \underline{\tilde{w}}_{r}  \underline{K}_{r}) \Big) \nonumber \\
&& - \dot{q} \sum_{r=1}^{N} f_r \tilde{\overline{w}}_{r} \nonumber \\
&& + (1 - q ) \sum_{r=1}^{N} \Big( \dot{f}_r \tilde{\overline{w}}_{r} + f_r ( - \tilde{\overline{w}}_{r} \overline{K}_{r} + \tilde{\overline{w}}_{r} \sum_{r=1}^{N} \tilde{\overline{w}}_{r}  \overline{K}_{r}) \Big) \nonumber \\
& = & \dot{q} \sum_{r=1}^{N} f_r \tilde{\underline{w}}_{r} + q \sum_{r=1}^{N} \Big( \dot{f}_r \tilde{\underline{w}}_{r} - I \alpha \textrm{sgn}\left(e \right) f_r ( \underline{\tilde{w}}_{r} - \underline{\tilde{w}}_{r} \sum_{r=1}^{N} \underline{\tilde{w}}_{r} ) \Big) \nonumber \\
&& - \dot{q} \sum_{r=1}^{N} f_r \tilde{\overline{w}}_{r} \nonumber \\
&& + (1 - q ) \sum_{r=1}^{N} \Big( \dot{f}_r \tilde{\overline{w}}_{r} - I \alpha \textrm{sgn}\left(e \right)  f_r ( \tilde{\overline{w}}_{r} - \tilde{\overline{w}}_{r} \sum_{r=1}^{N} \tilde{\overline{w}}_{r} ) \Big)
\end{eqnarray}

The equation \eqref{sumwr} is correct by definition:
\begin{equation} \label{sumwr}
\sum_{r=1}^{N} \tilde{\underline{w}}_{r} =1 \;\;\text{and}\;\;\; \sum_{r=1}^{N} \tilde{\overline{w}}_{r} =1
\end{equation}

By using \eqref{ar}, \eqref{br},  \eqref{dotq} and \eqref{sumwr}, the following function can be achieved:
\begin{eqnarray} \label{dotyn3}
\dot{y}_N & = & - \frac{1}{F (\tilde{\underline{W}} - \tilde{\overline{W}})^T}  \alpha \textrm{sgn}\left(e \right) \sum_{r=1}^{N} f_r (\tilde{\underline{w}}_{r} - \tilde{\overline{w}}_{r} ) \nonumber \\
&& + \sum_{r=1}^{N} \dot{f}_r (q \tilde{\underline{w}}_{r} + (1-q) \tilde{\overline{w}}_{r} ) \nonumber \\
 & = & - \alpha \textrm{sgn}\left(e \right) \nonumber \\
&& + \sum_{r=1}^{N} \Big[ \Big( \sum_{i=1}^{I} (\dot{a}_{ri} x_i + a_{ri} \dot{x}_i ) + \dot{b}_r \Big) \nonumber \\
&&(q \tilde{\underline{w}}_{r} + (1-q) \tilde{\overline{w}}_{r} ) \Big]
\end{eqnarray}

If \eqref{dotyn3} is inserted into the candidate Lyapunov function, \eqref{dotV2} can be obtained:
\begin{eqnarray} \label{dotV2}
\dot{V} & = & \dot{e} e +\frac{1}{\gamma}\dot{\alpha}(\alpha-\alpha^*) = e (\dot{y}_N - \dot{y}) +\frac{1}{\gamma}\dot{\alpha}(\alpha-\alpha^*)  \nonumber \\
& = & e \bigg[ - \alpha \textrm{sgn}\left(e \right) + \sum_{r=1}^{N} \Big[ \Big( \sum_{i=1}^{I} (\dot{a}_{ri} x_i + a_{ri} \dot{x}_i ) + \dot{b}_r \Big) \nonumber \\
&& (q \tilde{\underline{w}}_{r} + (1-q) \tilde{\overline{w}}_{r} ) \Big] - \dot{y} \bigg] +\frac{1}{\gamma}\dot{\alpha}(\alpha-\alpha^*)  \nonumber \\
\end{eqnarray}
\begin{eqnarray*}
& =&  e \bigg[ - \alpha \textrm{sgn}\left(e \right) + \sum_{r=1}^{N} \Big[ \Big( \sum_{i=1}^{I} \big(-  (x_{i} \alpha \textrm{sgn}\left(e \right) \nonumber \\
& & \frac{(q \tilde{\underline{w}}_{r} + (1-q) \tilde{\overline{w}}_{r} )}{(q \tilde{\underline{w}}_{r} + (1-q) \tilde{\overline{w}}_{r} )^{T} (q \tilde{\underline{w}}_{r} + (1-q) \tilde{\overline{w}}_{r} ) } ) x_i + a_{ri} \dot{x}_i \big) \nonumber\\
&& - \alpha \textrm{sgn}\left(e \right) \frac{(q \tilde{\underline{w}}_{r} + (1-q) \tilde{\overline{w}}_{r} )}{(q \tilde{\underline{w}}_{r} + (1-q) \tilde{\overline{w}}_{r} )^{T} (q \tilde{\underline{w}}_{r} + (1-q) \tilde{\overline{w}}_{r} ) } \Big)\nonumber \\
&&(q \tilde{\underline{w}}_{r} + (1-q) \tilde{\overline{w}}_{r} ) \Big] - \dot{y} \bigg] +\frac{1}{\gamma}\dot{\alpha}(\alpha-\alpha^*)    \nonumber \\
\end{eqnarray*}
\begin{eqnarray*}
& =&  e \bigg[ - \alpha \textrm{sgn}\left(e \right)  \nonumber \\
&& + \sum_{r=1}^{N} \Big[  \sum_{i=1}^{I} \big(-  \alpha \textrm{sgn}\left(e \right)  x^2_i + a_{ri} \dot{x}_i (q \tilde{\underline{w}}_{r} + (1-q) \tilde{\overline{w}}_{r} ) \big) \nonumber\\
&& - \alpha \textrm{sgn}\left(e \right) \Big] - \dot{y} \bigg] +\frac{1}{\gamma}\dot{\alpha}(\alpha-\alpha^*) \nonumber \\
\end{eqnarray*}
\begin{eqnarray*}
& =&  e \bigg[ - 2 \alpha \textrm{sgn}\left(e \right)+ \sum_{r=1}^{N} \Big[ \sum_{i=1}^{I} \big(-  \alpha \textrm{sgn}\left(e \right)  x^2_i \nonumber \\
&+& a_{ri} \dot{x}_i (q \tilde{\underline{w}}_{r} + (1-q) \tilde{\overline{w}}_{r} ) \big) \Big] - \dot{y} \bigg] +\frac{1}{\gamma}\dot{\alpha}(\alpha-\alpha^*) \nonumber \\
\end{eqnarray*}
\begin{eqnarray*}
& =&  - \mid e \mid 2 \alpha + e \bigg[ \sum_{r=1}^{N} \Big[ \sum_{i=1}^{I} \big(-  \alpha \textrm{sgn}\left(e \right)  x^2_i
\nonumber \\ &+& a_{ri} \dot{x}_i (q \tilde{\underline{w}}_{r} + (1-q) \tilde{\overline{w}}_{r} ) \big) \Big] - \dot{y} \bigg] +\frac{1}{\gamma}\dot{\alpha}(\alpha-\alpha^*) \nonumber \\
\end{eqnarray*}
\begin{eqnarray*}
\dot{V} & < &  - \mid e \mid  2 \alpha  + \mid e \mid \bigg[ \sum_{r=1}^{N} \Big[ \sum_{i=1}^{I} \big( - \alpha B_{x^2} \nonumber \\
&+& B_{a} B_{\dot{x}} (q \tilde{\underline{w}}_{r} + (1-q) \tilde{\overline{w}}_{r} ) \big)  \Big] + B_{\dot{y}} \bigg] +\frac{1}{\gamma}\dot{\alpha}(\alpha-\alpha^*) \nonumber \\
\end{eqnarray*}
\begin{eqnarray*}
& < &  - \mid e \mid 2 \alpha  + \mid e \mid \bigg[ \sum_{r=1}^{N} \Big[  -I \alpha B_{x^2} + I B_{a} B_{\dot{x}} (q \tilde{\underline{w}}_{r} \nonumber \\
\end{eqnarray*}
\begin{eqnarray*}
&+& (1-q) \tilde{\overline{w}}_{r} )  \Big] + B_{\dot{y}} \bigg] +\frac{1}{\gamma}\dot{\alpha}(\alpha-\alpha^*) \nonumber \\
& < &   - \mid e \mid  2 \alpha  + \mid e \mid \bigg[  -I \alpha B_{x^2} + I B_{a} B_{\dot{x}} (q \sum_{r=1}^{N} \tilde{\underline{w}}_{r} \nonumber \\
&+& (1-q) \sum_{r=1}^{N} \tilde{\overline{w}}_{r} )  + B_{\dot{y}} \bigg] +\frac{1}{\gamma}\dot{\alpha}(\alpha-\alpha^*) \nonumber \\
\end{eqnarray*}
\begin{eqnarray*}
& < &  - \mid e \mid  2 \alpha  + \mid e \mid \bigg[ -I \alpha B_{x^2} + I B_{a} B_{\dot{x}} + B_{\dot{y}} \bigg]  \nonumber \\ &+&
\frac{1}{\gamma}\dot{\alpha}(\alpha-\alpha^*) \nonumber \\
\end{eqnarray*}
\begin{eqnarray*}
& < &  - \mid e \mid \Big( 2 \alpha + I \alpha B_{x^2} \Big) + \mid e \mid \bigg[ I B_{a} B_{\dot{x}}  + B_{\dot{y}} \bigg]  \nonumber \\
&+&\frac{1}{\gamma}\dot{\alpha}(\alpha-\alpha^*) \nonumber \\
\end{eqnarray*}
where
\begin{equation*}
\alpha^*\geq\ \frac{ 2 (I B_{a} B_{\dot{x}}  + B_{\dot{y}} ) } {2+I B_{x^2}}
\end{equation*}
and $\alpha^*$ is considered to be an unknown parameter which is determined during the adaptation of the learning rate. This is an obvious superiority of the current approach over past approaches in which the upper bounds of the states of the system should be known a priori.
\begin{eqnarray}
\dot{V} & \leq & -\mid e\mid ( 2 \alpha + I \alpha B_{x^2} ) +\mid e\mid (  I B_{a} B_{\dot{x}}  + B_{\dot{y}} ) \nonumber \\
& + & (2+I B_{x^2}) \alpha^*\mid e\mid - (2+I B_{x^2}) \alpha^*\mid e\mid \nonumber \\
& + & \frac{1}{\gamma}(\alpha-\alpha^*)\dot{\alpha}
\end{eqnarray}

\begin{eqnarray}
\dot{V} & \leq & \mid e\mid ( I B_{a} B_{\dot{x}}  + B_{\dot{y}} ) - (2+I B_{x^2}) \alpha^* \mid e\mid \nonumber \\
&& -(2+I B_{x^2}) (\alpha^*-\alpha) \mid e \mid + \frac{1}{\gamma}(\alpha-\alpha^*)\dot{\alpha}
\end{eqnarray}
and further:
\begin{eqnarray}
\dot{V} & \leq & \mid e\mid ( I B_{a} B_{\dot{x}}  + B_{\dot{y}} ) - (2+I B_{x^2}) \alpha^* \mid e\mid \nonumber \\
&&+(\alpha^*-\alpha) \Big[ (2+I B_{x^2}) \mid e \mid - \frac{1}{\gamma}\dot{\alpha} \Big]
\end{eqnarray}
using the adaptation law for the adaptive learning rate ($\alpha$) as:
\begin{eqnarray}
\dot \alpha= (2+I B_{x^2}) \gamma \mid e\mid-\nu\gamma \alpha
\end{eqnarray}
in which $\nu$ has a small real value. Using this adaptation law, the time derivative of the Lyapunov function can be rewritten as:
\begin{eqnarray}
\dot{V} & \leq & \mid e\mid(I B_{a}B_{\dot{x}}+B_{\dot{y}}) \\ \nonumber
&-& (2+I B_{x^2}) \alpha^* \mid e\mid +( \alpha^*-\alpha)\nu\alpha
\end{eqnarray}
so that:
\begin{eqnarray}
\dot{V} & \leq & \mid e\mid ( I B_{a} B_{\dot{x}}  + B_{\dot{y}} ) \\ \nonumber
&-& (2+I B_{x^2}) \alpha^* \mid e\mid -\nu( \alpha-\frac{\alpha^*}{2})^2+\frac{\nu\alpha^{*2}}{4}
\end{eqnarray}
considering the fact that $\alpha^*\geq\ \frac{ 2 (I B_{a} B_{\dot{x}}  + B_{\dot{y}} ) } {2+I B_{x^2}}$ we have: $\mid e\mid ( I B_{a} B_{\dot{x}}  + B_{\dot{y}} ) - \frac{\alpha^*}{2} ( 2+I B_{x^2} ) \mid e\mid\leq 0$ and consequently:
\begin{eqnarray}
\dot{V} \leq -\frac{\alpha^*}{2}( 2+I B_{x^2})\mid e \mid +\frac{\nu\alpha^{*2}}{4}
\end{eqnarray}
So that error converges to a very small region around zero in which $\mid e \mid\leq\frac{\alpha^*\nu}{ 2 ( 2+I B_{x^2} )}$ and it remains there. It should also be noted that $\nu$ is a small user defined positive number which can be selected as small as desired to make this neighborhood as narrow as requested by the user.  
\bibliography{ieee-ie-bib-file}

\begin{thebibliography}{10}
\providecommand{\url}[1]{#1}
\csname url@samestyle\endcsname
\providecommand{\newblock}{\relax}
\providecommand{\bibinfo}[2]{#2}
\providecommand{\BIBentrySTDinterwordspacing}{\spaceskip=0pt\relax}
\providecommand{\BIBentryALTinterwordstretchfactor}{4}
\providecommand{\BIBentryALTinterwordspacing}{\spaceskip=\fontdimen2\font plus
\BIBentryALTinterwordstretchfactor\fontdimen3\font minus
  \fontdimen4\font\relax}
\providecommand{\BIBforeignlanguage}[2]{{%
\expandafter\ifx\csname l@#1\endcsname\relax
\typeout{** WARNING: IEEEtran.bst: No hyphenation pattern has been}%
\typeout{** loaded for the language `#1'. Using the pattern for}%
\typeout{** the default language instead.}%
\else
\language=\csname l@#1\endcsname
\fi
#2}}
\providecommand{\BIBdecl}{\relax}
\BIBdecl

\bibitem{1321083}
R.-J. Wai and P.-C. Chen, ``Intelligent tracking control for robot manipulator
  including actuator dynamics via tsk-type fuzzy neural network,'' \emph{IEEE
  Trans. Fuzzy Syst.}, vol.~12, no.~4, pp. 552--560, 2004.

\bibitem{5575419}
C.-H. Lu, ``Wavelet fuzzy neural networks for identification and predictive
  control of dynamic systems,'' \emph{IEEE Trans. Ind. Electron.}, vol.~58,
  no.~7, pp. 3046--3058, 2011.

\bibitem{4813268}
Y.-X. Liao, J.~hua She, and M.~Wu, ``Integrated hybrid-pso and fuzzy-nn
  decoupling control for temperature of reheating furnace,'' \emph{IEEE Trans.
  Ind. Electron.}, vol.~56, no.~7, pp. 2704--2714, 2009.

\bibitem{1593641}
C.-T. Lin, C.-M. Yeh, S.-F. Liang, J.-F. Chung, and N.~Kumar,
  ``Support-vector-based fuzzy neural network for pattern classification,''
  \emph{IEEE Trans. Fuzzy Syst.}, vol.~14, no.~1, pp. 31--41, 2006.

\bibitem{4292189}
I.-H. Li, W.-Y. Wang, S.-F. Su, and Y.-S. Lee, ``A merged fuzzy neural network
  and its applications in battery state-of-charge estimation,'' \emph{IEEE
  Trans. Energy Convers.}, vol.~22, no.~3, pp. 697--708, 2007.

\bibitem{1996}
L.~C.-T. and L.~C.~S. G., \emph{Neural Fuzzy Systems: A Neuro-Fuzzy Synergism
  to Intelligent Systems}.\hskip 1em plus 0.5em minus 0.4em\relax Upper Saddle
  River, NJ: Prentice Hall, 1996.

\bibitem{Mendel_kitap}
J.~M. Mendel, \emph{Uncertain Rule-Based Fuzzy Logic System: Introduction and
  New Directions}.\hskip 1em plus 0.5em minus 0.4em\relax {P}rentice {H}all,
  {U}pper {S}addle {R}iver, 2001.

\bibitem{chia09}
C.-F. Juang and C.-H. Hsu, ``Reinforcement interval type-2 fuzzy controller
  design by online rule generation and q-value-aided ant colony optimization,''
  \emph{IEEE Trans. Syst. Man, Cybern. B, Cybern.}, vol.~39, no.~6, pp. 1528
  --1542, dec. 2009.

\bibitem{Juang1}
------, ``Reinforcement ant optimized fuzzy controller for mobile-robot
  wall-following control,'' \emph{IEEE Trans. Ind. Electron.}, vol.~56, no.~10,
  pp. 3931 --3940, oct. 2009.

\bibitem{6469210}
Y.-Y. Lin, J.-Y. Chang, and C.-T. Lin, ``A tsk-type-based self-evolving
  compensatory interval type-2 fuzzy neural network (tscit2fnn) and its
  applications,'' \emph{IEEE Trans. Ind. Electron.}, vol.~61, no.~1, pp.
  447--459, Jan 2014.

\bibitem{Astrom_Witternmark}
K.~Astrom and B.~Wittenmark, \emph{Adaptive Control}.\hskip 1em plus 0.5em
  minus 0.4em\relax Addison-Wesley, 1995.

\bibitem{Venelinov_1}
A.~Topalov and O.~Kaynak, ``Online learning in adaptive neurocontrol schemes
  with a sliding mode algorithm,'' \emph{IEEE Trans. Syst. Man, Cybern. B,
  Cybern.}, vol.~31, no.~3, pp. 445--450, december 2001.

\bibitem{5949558}
M.~Khanesar, E.~Kayacan, M.~Teshnehlab, and O.~Kaynak, ``Levenberg marquardt
  algorithm for the training of type-2 fuzzy neuro systems with a novel type-2
  fuzzy membership function,'' in \emph{Advances in Type-2 Fuzzy Logic Systems
  (T2FUZZ), 2011 IEEE Symposium on}, 2011, pp. 88--93.

\bibitem{Venelinov_2}
A.~Topalov, K.-C. Kim, J.-H. Kim, and B.-K. Lee, ``Fast genetic on-line
  learning algorithm for neural network and its application to temperature
  control,'' in \emph{IEEE Int. Conf. Evolutionary Computation, Nagoya, Japan},
  May 1996, pp. 649 --654.

\bibitem{5611774}
M.~Khanesar, M.~Teshnehlab, E.~Kayacan, and O.~Kaynak, ``A novel type-2 fuzzy
  membership function: application to the prediction of noisy data,'' in
  \emph{Computational Intelligence for Measurement Systems and Applications
  (CIMSA), 2010 IEEE International Conference on}, 2010, pp. 128--133.

\bibitem{Parma}
G.~Parma, B.~Menezes, and A.~Braga, ``Sliding mode algorithm for training
  multilayer artificial neural networks,'' \emph{Electronics Letters}, vol.~34,
  no.~1, pp. 97 --98, 1998.

\bibitem{Yu}
Y.~Shuanghe, Y.~Xinghuo, and M.~Zhihong, ``A fuzzy neural network approximator
  with fast terminal sliding mode and its applications,'' \emph{Fuzzy Sets and
  Systems}, vol. 148, no.~3, pp. 469--486, 11 2004.

\bibitem{Cascella}
G.~Cascella, F.~Cupertino, A.~Topalov, O.~Kaynak, and V.~Giordano, ``Adaptive
  control of electric drives using sliding-mode learning neural networks,'' in
  \emph{Industrial Electronics, 2005. ISIE 2005. Proceedings of the IEEE
  International Symposium on}, vol.~1, 2005, pp. 125 -- 130.

\bibitem{Efe2000}
M.~O. Efe, O.~Kaynak, and X.~Yu, ``Sliding mode control of a three degrees of
  freedom anthropoid robot by driving the controller parameters to an
  equivalent regime,'' \emph{ASME Journal of Dynamic Systems, Measurement, and
  Control}, vol. 122, no.~4, pp. 632 --640, December 2000.

\bibitem{6117080}
E.~Kayacan, O.~Cigdem, and O.~Kaynak, ``Sliding mode control approach for
  online learning as applied to type-2 fuzzy neural networks and its
  experimental evaluation,'' \emph{IEEE Trans. Ind. Electron.}, vol.~59, no.~9,
  pp. 3510--3520, 2012.

\bibitem{ACS:ACS1292}
E.~Kayacan and O.~Kaynak, ``Sliding mode control theory-based algorithm for
  online learning in type-2 fuzzy neural networks: application to velocity
  control of an electro hydraulic servo system,'' \emph{International Journal
  of Adaptive Control and Signal Processing}, vol.~26, no.~7, pp. 645--659,
  2012.

\bibitem{Begian}
M.~Begian, W.~Melek, and J.~Mendel, ``Parametric design of stable type-2 tsk
  fuzzy systems,'' in \emph{Fuzzy Information Processing Society, 2008. NAFIPS
  2008. Annual Meeting of the North American}, May 2008, pp. 1--6.

\bibitem{Slotine}
J.-J. Slotine and W.~Li, \emph{Applied Nonlinear Control}.\hskip 1em plus 0.5em
  minus 0.4em\relax Prentice Hall, 1991.

\bibitem{Topalovv}
S.~Ahmed, N.~Shakev, A.~Topalov, K.~Shiev, and O.~Kaynak, ``Sliding mode
  incremental learning algorithm for interval type-2 takagi-sugeno-kang fuzzy
  neural networks,'' \emph{Evolving Systems}, vol.~3, no.~3, pp. 179 --188,
  September 2012.

\bibitem{363441}
C.-C. Ku and K.~Lee, ``Diagonal recurrent neural networks for dynamic systems
  control,'' \emph{IEEE Trans. Neural Netw.}, vol.~6, no.~1, pp. 144--156,
  1995.

\bibitem{Abiyev1}
R.~Abiyev and O.~Kaynak, ``Type 2 fuzzy neural structure for identification and
  control of time-varying plants,'' \emph{IEEE Trans. Ind. Electron.}, vol.~57,
  no.~12, pp. 4147--4159, Dec 2010.

\bibitem{Khanesar}
M.~Khanesar, E.~Kayacan, M.~Teshnehlab, and O.~Kaynak, ``Analysis of the noise
  reduction property of type-2 fuzzy logic systems using a novel type-2
  membership function,'' \emph{IEEE Trans. Syst. Man, Cybern. B, Cybern.},
  vol.~41, no.~5, pp. 1395--1406, Oct 2011.

\bibitem{khanesar2007hybrid}
M.~Khanesar, M.~Shoorehdeli, and M.~Teshnehlab, ``Hybrid training of recurrent
  fuzzy neural network model,'' in \emph{Mechatronics and Automation, 2007.
  ICMA 2007. International Conference on}.\hskip 1em plus 0.5em minus
  0.4em\relax IEEE, 2007, pp. 2598--2603.

\end{thebibliography}
\bibliographystyle{IEEEtran}

\begin{IEEEbiography}[{\includegraphics[width=1in,height=1.25in,clip,keepaspectratio]{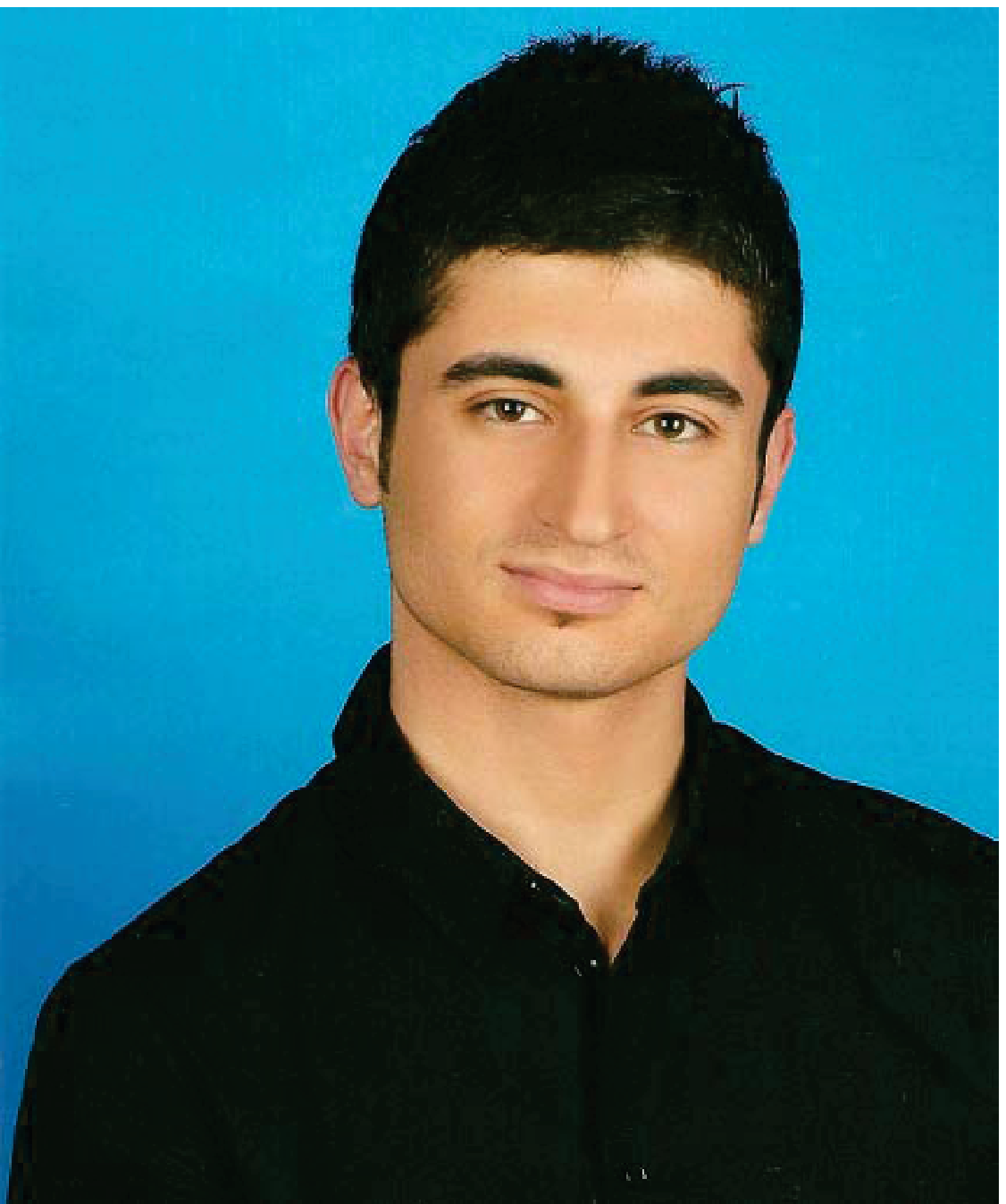}}]{Erkan Kayacan} (S\textquoteright 12) was born in Istanbul, Turkey, on April 17, 1985. He received the B.Sc. and the M.Sc. degrees in mechanical engineering from Istanbul Technical University, Istanbul, in 2008 and 2010, respectively. He is currently pursuing the Ph.D. degree with the Division of Mechatronics, Biostatistics and Sensors (MeBioS), KU Leuven, Leuven, Belgium.

He is a Research Assistant at the University of Leuven. His current research interests include model predictive control, moving horizon estimation, decentralized and distributed control, intelligent control and mechatronics.
\end{IEEEbiography}

\begin{IEEEbiography}[{\includegraphics[width=1in,height=1.25in,clip,keepaspectratio]{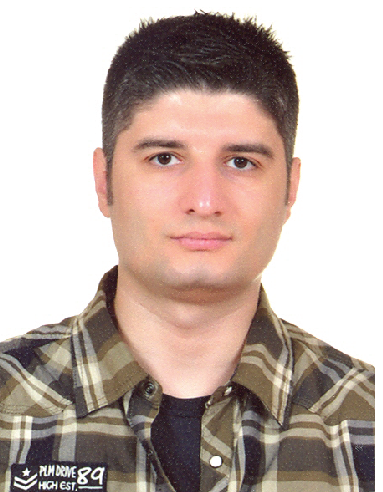}}]{Erdal Kayacan} (S\textquoteright 06-SM\textquoteright 12)  was born in Istanbul, Turkey on January 7, 1980. He received a B.Sc. degree in electrical engineering from in 2003 from Istanbul Technical University in Istanbul, Turkey as well as a M.Sc. degree in systems and control engineering in 2006 from Bogazici University in Istanbul, Turkey. In September 2011, he received a Ph.D. degree in electrical and electronic engineering at Bogazici University in Istanbul, Turkey. After finishing his post-doctoral research in KU Leuven at the division of mechatronics, biostatistics and sensors (MeBioS), he is currently pursuing his research in Nanyang Technological University at the School of Mechanical and Aerospace Engineering as an assistant professor. His research areas are unmanned aerial vehicles, robotics, mechatronics, soft computing methods and model predictive control.

\end{IEEEbiography}

\begin{IEEEbiography}[{\includegraphics[width=1in,height=1.25in,clip,keepaspectratio]{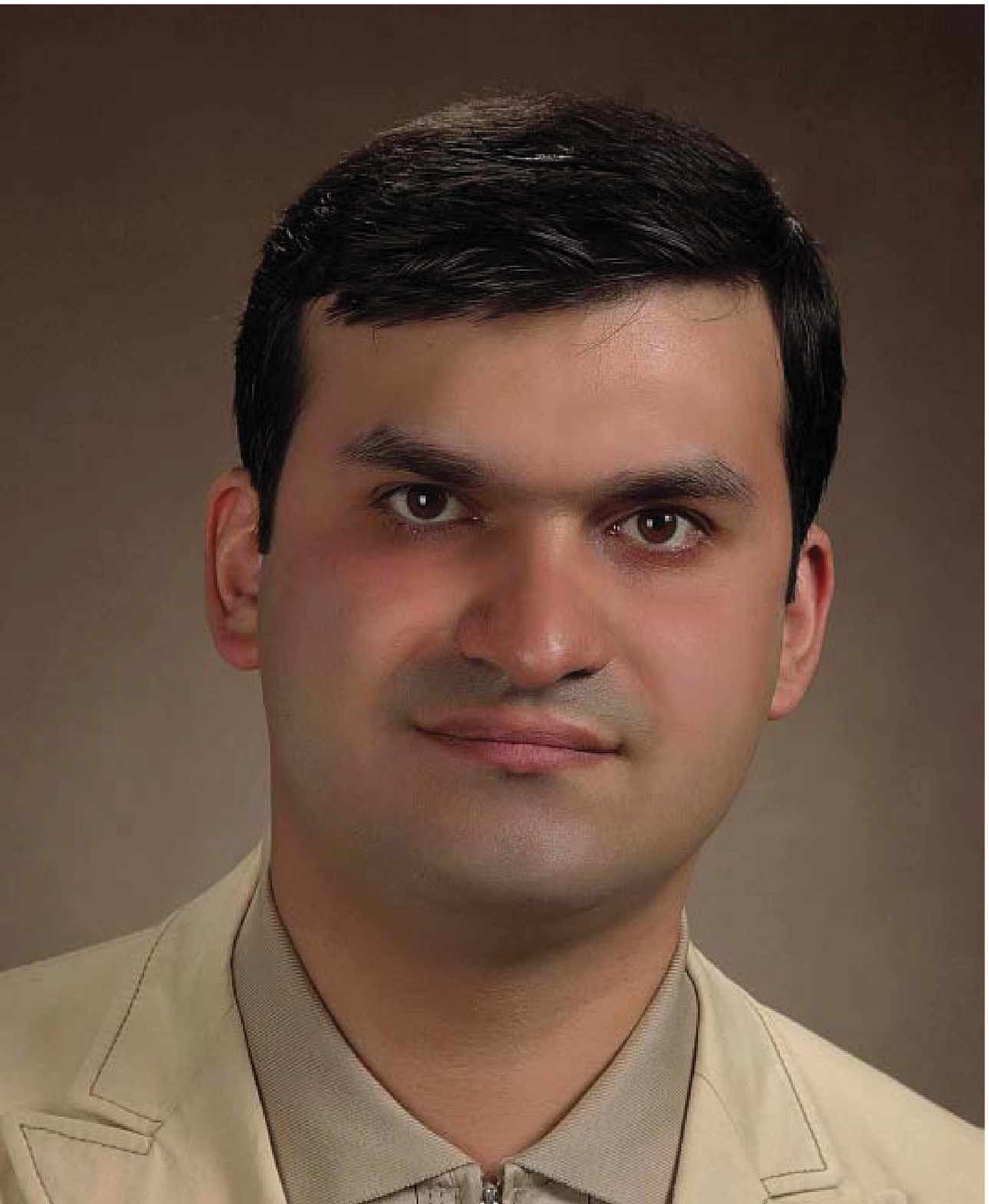}}]{Mojtaba Ahmadieh Khanesar } (S\textquoteright 07-M\textquoteright 12) received the M.S. and Ph.D. degree in Control Engineering from K. N. Toosi University of Technology, Tehran, Iran, in 2007 and 2012 respectively. In 2010, he has held a 9 months visiting student position at Bogazici University, Istanbul, Turkey. He is currently an assistant professor in the Control and Electrical Engineering Department, Semnan University, Semnan, Iran. His current research interests are identification, mechatronics, sliding mode control, adaptive controller design, fuzzy systems, intelligent optimization and model predictive control.

Dr. Khanesar is a member of IEEE technical committee on soft computing and IEEE technical committee on networked control systems. 
\end{IEEEbiography}

\end{document}